\documentclass[%
reprint,
superscriptaddress,
preprintnumbers,
nofootinbib,
amsmath,
amssymb,
aps,
prc,
]{revtex4-2}

\usepackage{graphicx}
\usepackage{dcolumn}
\usepackage{hyperref}
\usepackage{microtype}
\usepackage{subcaption}
\begin{document}

\preprint{HUPD-2603}

\title{Thermal dileptons to probe the baryon-rich QCD matter in the forward region of LHC energy heavy-ion collisions}


\author{Motomi Oya}
\email{oya@quark.hiroshima-u.ac.jp}
\affiliation{Physics Program, Hiroshima University, Higashi-Hiroshima, 739-8526, Japan}
\author{Nicholas J. Benoit}
\email{njbenoit@hiroshima-u.ac.jp}
\affiliation{Physics Program, Hiroshima University, Higashi-Hiroshima, 739-8526, Japan}
\affiliation{Institute of Physics, Academia Sinica, Taipei, 11529, Taiwan}
\author{Chiho Nonaka}
\affiliation{Physics Program, Hiroshima University, Higashi-Hiroshima, 739-8526, Japan}
\affiliation{International Institute for Sustainability with Knotted Chiral Meta Matter, Hiroshima University, Higashi-Hiroshima 739-8511, Japan}
 \affiliation{Kobayashi Maskawa Institute, Nagoya University, Nagoya 464-8602, Japan.}
\author{Azumi Sakai}
\affiliation{Department of Applied Information Technology, Nagasaki Institute of Applied Science, Nagasaki 851-0193, Japan}
\author{Yorito Yamaguchi}
\affiliation{Physics Program, Hiroshima University, Higashi-Hiroshima, 739-8526, Japan}



\date{\today}

\begin{abstract}
We investigate thermal dilepton production from a quark–gluon plasma (QGP) with finite baryon 
chemical potential ($\mu_{\text{B}}$) in central Pb–Pb collisions at $\sqrt{s_{\text{NN}}}=5.02~\text{TeV}$. 
Recent studies suggest that sizable baryon densities can be achieved at forward rapidity even at LHC energies. 
We incorporate finite  $\mu_{\text{B}}$ into a (3+1)-dimensional hydrodynamic framework and find that $\mu_{\text{B}}$ exceeds 
500 MeV around $\eta_\text{s} = 6$ during the medium evolution.
Using this framework, we calculate thermal dilepton spectra over a wide rapidity range and 
evaluate the impact of finite $\mu_{\text{B}}$ on dilepton production. A suppression of 3–4~\% is 
observed in the forward-rapidity region 
$5.2 < y < 7.2$ due to the reduced quark–antiquark abundance at finite baryon density. 
We further examine the effective temperature extracted from dilepton mass spectra in the intermediate-mass region $1.2 < M_{\ell \ell} < 2.6~\text{GeV}$ . 
The effective temperature remains strongly correlated with the underlying hydrodynamic temperature and retains sensitivity 
to the early high-temperature stage of the QGP evolution. 
These results demonstrate that forward-rapidity dileptons remain effective thermometers while providing sensitivity to finite baryon density at the LHC.
\end{abstract}

\maketitle

\section{INTRODUCTION}\label{introduction}
Dileptons are unique probes of the thermodynamic properties of the quark-gluon plasma (QGP) because they are transparent to the strong interaction. 
For example, their invariant mass spectrum is the most reliable way to determine the initial QGP temperature~\cite{PhysRevLett.132.172301,physjc85}.
Temperatures from dileptons were first measured at the CERN-SPS experiments~\cite{specht20101,PhysRevLett.75.1272,helios3} followed by the RHIC~\cite{PhysRevC.81.034911} experiments.
Recently, the ALICE experiment at the LHC reported an excess of dielectrons in the low mass region ($0.2 < m_{e^+e^-} < 0.7~\text{GeV}$) at mid rapidity, but the large uncertainty prevents a clear temperature extraction~\cite{xl6m-vbqk}.
Understanding these experimental results requires a comparison with theoretical models describing the time evolution of the QGP, such as hydrodynamic models and transport models~\cite{LINNYK201650, 10.1142/S0217751X13400113, Kolb:2003dz}.

Conventionally model calculations assume a negligible baryon chemical potential.
Recently, this assumption was shown to breakdown in the forward rapidity region using the McLerran-Venugopalan model~\cite{PhysRevD.49.2233,PhysRevD.49.3352,PhysRevD.50.2225} to calculate the rapidity loss of nucleons~\cite{PhysRevC.99.014906}.
    In their calculation, interactions between participants and color flux tubes lead to a maximum baryon density of $15.8$~baryons/$\text{fm}^3$ in Pb-Pb collisions at $\sqrt{s_{\text{NN}}}=5.02~\text{TeV}$.
This was further supported by the McDIPPER model~\cite{PhysRevC.109.044916}  which calculates the rapidity loss of nucleons in a $k_{\perp}$-factorized color-glass-condensate (CGC) framework.
Moreover, Ref.~\cite{FUJII2026100289} simulated the space-time evolution of the baryon chemical potential in Pb-Pb collisions at LHC energy, showing that the maximum baryon chemical potential of particlization become $0.6~\text{GeV}$ in the forward region.
These studies indicate that the forward-rapidity region of Pb--Pb collisions at the LHC may reach baryon chemical 
potentials comparable to those encountered in low-energy heavy-ion collisions, while maintaining substantially 
higher temperatures. Such a high-temperature and finite-density QGP has not yet been systematically investigated using electromagnetic probes.

Effects of the finite baryon chemical potential on the dilepton production rate need to be assessed for better description of experimental data, as such a non-negligible baryon density is expected.
Finite baryon density suppresses  the thermal dilepton production rate through the imbalance between quarks and antiquarks~\cite{PhysRevLett.70.2860}.
 This suppression 
was also demonstrated in the next-to-LO calculation for intermediate masses, $1 < M_{\ell \ell} < 3\text{ GeV}$, of dileptons in $7.7~\text{GeV} < \sqrt{s_{\text{NN}}} < 200~\text{GeV}$ covered by the RHIC Beam Energy Scan (BES) programs~\cite{PhysRevC.109.044915}.
Unfortunately, after the hydrodynamic evolution the suppression did not reach a detectable level with the current experimental precision~\cite{PhysRevC.109.044915}.
However, the possible suppression in forward rapidities at LHC energies has not yet been explored.
This combined with the recent models that predict a finite baryon chemical potential motivates our work.

An important open question is whether thermal dileptons remain reliable thermometers in this forward-rapidity regime. 
The extraction of effective temperatures from dilepton mass spectra has been extensively discussed for nearly baryon-free matter. 
However, the presence of a sizable baryon chemical potential and strong rapidity dependence may 
alter both the dilepton yield and its interpretation in terms of the underlying hydrodynamic temperature.

In this work, we investigate thermal dileptons in Pb--Pb collisions at $\sqrt{s_{\text{NN}}}=5.02$  TeV over a wide rapidity range.  
Focusing on the intermediate-mass region, $1 < M_{\ell\ell} < 3$ GeV, where QGP radiation dominates over hadronic contributions, we evaluate the impact of finite baryon chemical potential on the dilepton mass spectrum using a hydrodynamic description of the medium. 
Here we use ideal hydrodynamics for a first exploratory study towards future viscous hydrodynamic calculations including finite baryon density and baryon diffusion.
We further examine the relation between dilepton effective temperatures and hydrodynamic temperatures in the forward-rapidity region, and assess the sensitivity of thermal dileptons to the high-temperature finite-density QGP predicted by recent baryon-stopping models.

This paper is organized as follows: Sec. \ref{two} describes our model of QGP space-time evolution based on the relativistic ideal hydrodynamics.
The initial conditions are defined by the Monte Carlo (MC) Glauber, and the formula for thermal dilepton production rate through quark-antiquark annihilation.
We newly introduce the baryon density to the calculations at all stages. 
In Sec. \ref{three}, our calculation results are shown to discuss the effects of the finite baryon chemical potential on the dilepton mass spectrum by a comparison of the results with and without baryon chemical potential.
Effective temperatures are extracted from calculated spectra, and we discuss the sensitivity to the initial temperature in $5.2 < y < 7.2$. 
Finally, our results and conclusions are summarized in Sec.~\ref{four}.

\section{THEORETICAL FRAMEWORK}\label{two}
We employ $(3+1)$D ideal hydrodynamics to describe the QGP space-time evolution.
The initial conditions are created by a tilted Monte Carlo (MC)-Glauber~\cite{PhysRevC.81.054902,PhysRevC.106.L061901} model.
It is tuned to reproduce the charged particle multiplicity measured by ALICE, and the initial baryon density predicted by McDIPPER~\cite{PhysRevC.109.044916}.
That acts as our background for the dilepton production rate, which was calculated using the Born approximation considering only LO processes.
Our calculations focus on the most central Pb-Pb collisions, where the dilepton rate is expected to be the most significant.

\subsection{Quark-gluon plasma space-time evolution}
We model the space-time evolution of the QGP using $(3+1)$D relativistic ideal hydrodynamics~\cite{HIRANO2006299,PhysRevC.75.014902,PhysRevC.83.021902,PhysRevC.84.011901,PhysRevC.82.041901,PhysRevC.81.054903,PhysRevC.83.034901,PhysRevC.86.024911}, which is extended with a finite baryon chemical potential.
Since this is the first hydrodynamical study of dilepton production with large $n_\text{B}$ at LHC energy, we adopt the ideal hydrodynamical framework to cleanly isolate the effect of $n_\text{B}$. Furthermore, 
viscous corrections for the thermal dilepton production in the mass range $1 < M_{\ell \ell}< 3\text{ GeV}$ are known to be relatively small, as calculated at RHIC energies~\cite{PhysRevC.101.044904}.
In ideal hydrodynamics, the conservation laws of energy, momentum, and total charge are
\begin{gather}\label{eq:totalcharge}
  \nabla_\mu J^\mu = 0,
  \\ \label{eq:energymomentum}
  \nabla_\mu T^{\mu \nu} = 0,
\end{gather}
where $J^\mu$ is the baryon charge current and $T^{\mu \nu}$ is the energy-momentum tensor.
They are written as
\begin{gather}\label{eq:hydro_current}
  J^{\mu} = n_{\text{B}} u^{\mu},
  \\ \label{eq:hydro_energymom}
  T^{\mu \nu} = (e+p) u^{\mu} u^{\nu} - p g^{\mu \nu},
\end{gather}
where $u^\mu$ is the fluid four-velocity, normalized as $u^\mu u_\mu$ =1.
The symbols $e$, $p$ and $n_\text{B}$ denote the energy density, the pressure and the baryon number density, respectively.

To solve Eqs.~\eqref{eq:totalcharge} and \eqref{eq:energymomentum} we adopt the equation of state (EoS) NEOS-4D~\cite{PhysRevC.110.044905}.
It smoothly connects lattice QCD in high temperatures to a hadron resonance gas model in low temperatures.
This EoS is valid  
in the temperature region $0 < T < 1$ GeV and for a baryon chemical potential of $-0.24 < \mu_{\text{B}} < 0.6$ GeV\footnote{Although, NEOS-4D also provides values for an electric chemical potential of $-0.05 < \mu_{\text{Q}} < 0.01$ GeV and strangeness chemical potential of $-0.1 < \mu_{\text{S}} < 0.25$ GeV, we do not require them for this work.}.
The corresponding EoS tables are used for the numerical calculations in this work~\cite{qcdneos4d}.

The hydrodynamics calculation is performed in Milne coordinates ($\tau$, $x$, $y$, $\eta_\text{s}$).
Here $x$ and $y$ are the same as the Minkowski coordinates, the proper time and the space-time rapidity are defined as $\tau = \sqrt{t^2 - z^2}$ and $\eta_\text{s} = (1/2)\ln{(t+z)/(t-z)}$, respectively.
We use a grid of $169 \times 169 \times 120$ in $-16.9 < x,y < 16.9$ fm, and $-12.0 < \eta_\text{s} < 12.0$.


The space-time evolution of fluid is calculated by solving the Riemann problem, where the initial condition is the discontinuities of adjacent cells.
Those discontinuous surfaces instantaneously develop into multiple non-linear shock waves~\cite{Mignone_2005}.
The postshock variables are calculated by solving the conservation laws, Eq.\eqref{eq:totalcharge} and Eq.\eqref{eq:energymomentum}, using the approximation scheme in Ref.~\cite{AKAMATSU201434}.
That scheme determines the postshock pressure using a single-variable Newton-Raphson algorithm.
It updates the candidate postshock pressure iteratively until the continuity of the flow velocity is satisfied.
That algorithm approximates low baryon density, which needed to be changed for this work.

We extend the algorithm to use a new two-variable Newton-Raphson algorithm to handle the finite baryon density dependence of $e = (p, n_\text{B})$ during the iteration.
The conserved variables are updated from the flow velocity, and the thermodynamic variables are subsequently derived from a second iterative calculation.
The partial derivatives of $e(p, n_\text{B})$ used in the iteration are given by NEOS-4D with finite $n_\text{B}$.

\subsection{Initial Energy Density}
We determine the initial energy density using the tilted MC-Glauber model~\cite{PhysRevC.81.054902,PhysRevC.106.L061901}. 
The initial energy density is described by the combination of the number density of participants $n_\text{w}(x,y)$ and the number density of binary collisions $n_\text{b.c.}(x,y)$.
Those two factors are weighted by a free parameter $\alpha$,
\begin{equation}
\label{eq:epsilon}
    e(x,y) = \frac{e_0}{\tau_0}\left[\alpha n_\text{b.c.}(x,y) + (1-\alpha) n_\text{w}(x,y)\right],
\end{equation}
where we use $\alpha = 0.05$, following Ref.~\cite{PhysRevC.108.064904}.
In Eq.~\eqref{eq:epsilon}, $e_0$ is the energy normalization factor and $\tau_0$ is the initial time of the hydrodynamic calculation.
$n_\text{b.c.}$ and $n_\text{w}$ are calculated from: the positions of the nucleons sampled from the Woods-Saxon distribution, the inelastic nucleon-nucleon cross-section $\sigma_{\text{NN}}$ from a fit of experimental data over a range of collision energies~\cite{102419-060007}, and the spatial profile of nucleons which are assumed to be Gaussian with a width of $D=\sigma_{\text{NN}}/8\pi$.
Parameters for the Woods-Saxon distribution are taken from Ref.~\cite{SHOU2015215}.
The number of binary collisions for Gaussian nucleons $i$ and $j$ in Eq.\eqref{eq:epsilon} is given by
\begin{equation}
n_\text{b.c.}(x,y)=\sum_{i,j\in \text{pairs}}\frac{1}{2\pi D}\exp\left(\frac{-\lvert \vec{r}-\vec{R_{ij}}\rvert^2}{2D}\right),
\end{equation}
where $\vec{r} = (x, y)$, and $\vec{R_{ij}} = \frac{1}{2}(\vec{r_i} + \vec{r_j})$ which specifies the collision point between the centers of the nucleons $i$ and $j$.

Next, Eq.~\eqref{eq:epsilon} can be extended 
into transverse plane and rapidity direction as
\begin{equation}
    e(x,y,\eta_\text{s}) = \frac{e_0}{\tau_0} [\alpha n_\text{b.c.}(x,y) + (1-\alpha)n_{\text{w}}(x,y,\eta_\text{s})]H(\eta_\text{s}),
\end{equation}
where 
\begin{equation}\label{eq:wounded}
n_{\text{w}}(x,y,\eta_\text{s}) = W_+(x,y)f_+(\eta_\text{s}) + W_-(x,y)f_-(\eta_\text{s}).
\end{equation}
$W_{+/-}(x,y)$ is the number of wounded nucleons defined as
\begin{equation}
W_{+/-}(x,y)=\sum_{i\in\text{wounded}}\sum_{j=1}^{N_{b,i}}\frac{1}{N_{b,i}}\frac{1}{2\pi D}\exp\left(\frac{-\lvert r-R_{ij}\rvert^2}{2D}\right),
\end{equation}
where $\sum_{i\in\text{wounded}}$ denotes the sum over all wounded nucleons, and $\sum_{j=1}^{N_{b,i}}$ is the sum over their collision partners.
$f_\pm(\eta_\text{s})$ are the asymmetric factors along the space-time rapidity given by
\begin{equation}
f_{\pm}(\eta_\text{s}) = 
\begin{cases}
    1, & \eta_\text{s} > \eta_{\text{m}}, \\
    \frac{\eta_{\text{m}} \pm \eta_\text{s}}{2\eta_{\text{m}}}, & -\eta_{\text{m}} \leq \eta_\text{s} \leq \eta_{\text{m}}, \\
    0, & \eta_\text{s} < -\eta_{m},
\end{cases}
\end{equation}
where $\eta_{\text{m}}$ defines the tilt of the fireball~\cite{PhysRevC.81.054902,PhysRevC.106.L061901}.
The longitudinal scaling factor $H(\eta_\text{s})$ is represented by
\begin{equation}
  H(\eta_\text{s}) = \exp\left[-\theta(|\eta_\text{s}| - \eta^0_{\text{s}}/2)\frac{(|\eta_\text{s}| - \eta^0_{\text{s}}/2)^2}{2{\sigma_\eta}^2}\right],
\end{equation}
where $\eta^0_{\text{s}}$ determines the width of the plateau and $\sigma_\eta$ characterizes the Gaussian tail~\cite{PhysRevC.81.054902,PhysRevC.106.L061901}.

\begin{figure}
	\centering 
	\includegraphics[width=\linewidth]{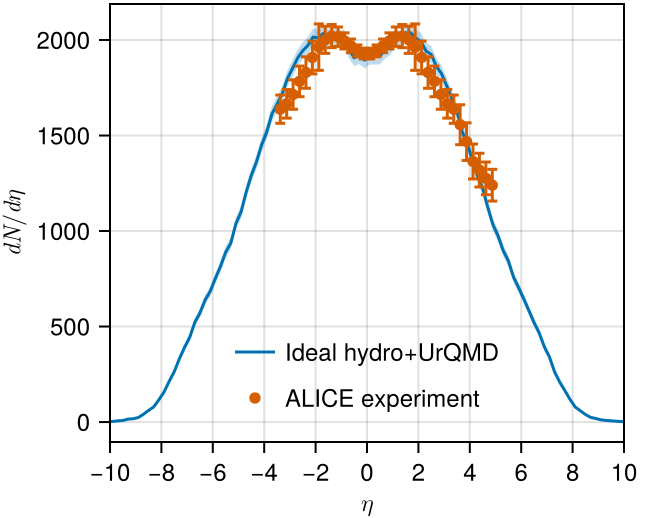}	
      \caption{Comparison of the charged particle multiplicity $dN_{\text{ch}}/d\eta$ between experimental data and the tuned simulation. The orange data points correspond to the ALICE experiment in 0-5 \% central Pb-Pb collisions at $\sqrt{s_{\text{NN}}} = 5.02\,\text{TeV}$~\cite{PhysRevLett.116.222302}. The blue solid line represents the result of our model.
      } 
	\label{eta_distribution}%
\end{figure}

We have tuned $e_0$, $\eta^0_{\text{s}}$, $\eta_{\text{m}}$ and  $\sigma_{\eta}$ to reproduce the charged particle multiplicity for 0-5\% Pb-Pb collisions at $\sqrt{s_{\text{NN}}} = 5.02~\text{TeV}$ measured by ALICE~\cite{PhysRevLett.116.222302}.
The hadron distribution is estimated by the following procedures.
\begin{enumerate}
    \item The switch from hydrodynamics to microscopic kinetic theory on the particlization hypersurface is defined by an energy density of $0.26\,\text{GeV}/\text{fm}^3$.
    \item The open-source tool iS3D~\cite{McNelis:2019auj,iS3D} samples hadrons from the hypersurface using the Cooper-Frye formula~\cite{PhysRevD.10.186}.
    \item Final state interactions between the hadrons are handled using UrQMD model~\cite{BASS1998255,Bleicher_1999} until kinetic freezeout.
\end{enumerate}
Figure~\ref{eta_distribution} illustrates that our model reproduces the charged particles multiplicity $dN_{\text{ch}}/d\eta$ for central collisions~\cite{PhysRevLett.116.222302}.
The impact parameter $b$ in the initial condition is fixed to $b = 0\,\text{fm}$. 

\subsection{Initial Baryon Density}
Similarly, we construct the initial baryon density considering the number of participants defined in Eq.\eqref{eq:wounded} as
\begin{equation}
  n_\text{B}(x,y,\eta_\text{s}) = \frac{n_0}{\tau_0} n_\text{w}(x,y,\eta_\text{s}) H(\eta_\text{s}) H_{\text{B}}(\eta_\text{s}),
\end{equation}
with the baryon normalization factor $n_0$.
The additional factor $H_\text{B}$ describes the longitudinal shape of the baryon distribution which has peaks in $\eta_\text{s} = \pm \eta_{\text{B}}$~\cite{PhysRevC.106.L061901} and is given by
\begin{equation}\label{eq:initialB}
  H_{\text{B}}(\eta_\text{s}) = \exp\left[ -\frac{(\eta_\text{s} - \eta_{\text{B}})^2}{2\sigma_{\text{B}}^2} \right] + \exp\left[ -\frac{(\eta_\text{s} + \eta_{\text{B}})^2}{2\sigma_{\text{B}}^2} \right],
\end{equation}
where $\eta_{\text{B}}$ and $\sigma_{\text{B}}$ define the longitudinal shape of the baryon distribution.

\begin{figure}[htbp]
  \centering
  \includegraphics[width=\linewidth]{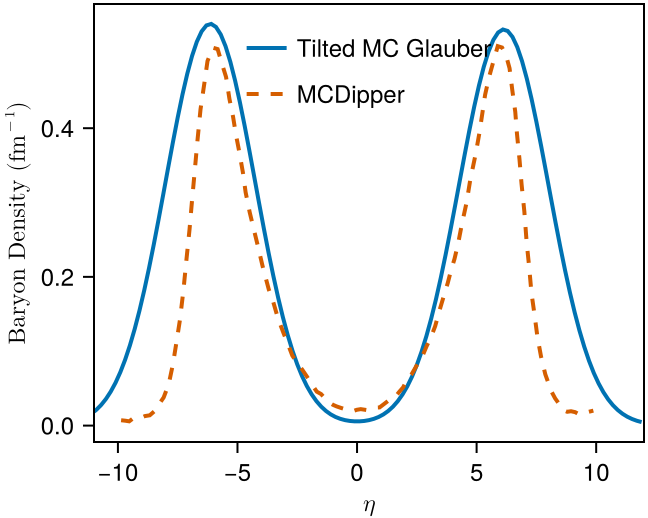}
  \caption{Comparison of the initial baryon density ($\text{fm}^{-1}$) between the McDIPPER prediction~\cite{PhysRevC.109.044916} and the tuned simulation.}
  \label{fig:baryon_distribution}
\end{figure}

Currently, the rapidity-dependent anti-baryon to baryon ratio has not been measured at $\sqrt{s_{\text{NN}}} = 5.02$ TeV.
Instead, we used the open-source McDIPPER framework~\cite{PhysRevC.109.044916} to fix $n_0$, $\eta_{\text{B}}$, and $\sigma_{\text{B}}$ in Eq.\eqref{eq:initialB}.
Figure~\ref{fig:baryon_distribution} shows the comparison between our initial baryon density and McDIPPER.
Although we do not perfectly reproduce the McDIPPER results due to the limited free parameters, it was tuned to start as close as possible.
Nevertheless, the outer part beyond $\eta \sim 7$ is excluded from the discussion in this study due to the large deviations from the McDIPPER.
Figure \ref{eta_distribution} incorporate the baryon number density shown in Figure \ref{fig:baryon_distribution}; however, we have confirmed that its impact on the charged particle multiplicity is negligible.
All the initial  condition parameters are listed in Table \ref{tab:initial_param}.
\begin{table}[h]
    \centering
    \caption{Initial parameters tuned to the central Pb-Pb collisions at $\sqrt{s_{\text{NN}}} = 5.02\,\text{TeV}$.}
    \label{tab:initial_param}
    \begin{tabular}{|cc|cc|cc|}
        \hline
        $\tau_0$   & $0.6\,\text{fm}$  & $e_0$           & $150\,\text{GeV}/\text{fm}^3$ & $n_0$      & $290\,\text{GeV}/\text{fm}^3$ \\
        $\eta_{\text{m}}$ & $3.36\,\text{fm}$ & $\eta^0_{\text{s}}$      & $2.2\,\text{fm}$                & $\eta_{\text{B}}$   & $14.0\,\text{fm}$ \\
                   &                       & $\sigma_{\eta}$ & $2.4\,\text{fm}$              & $\sigma_{\text{B}}$ & $3.0\,\text{fm}$ \\
        \hline
    \end{tabular}
\end{table}

\subsection{Thermal dilepton production rate}
We calculate the thermal dilepton production by the leading order (LO) $q\bar{q}$ annihilation $q+\bar{q} \to \ell^+\ell^-$ process.
Even though the next-to-LO results exist~\cite{PhysRevC.109.044915} the LO process is dominant within our considered invariant mass range $1 < M_{\ell \ell} < 3\text{ GeV}$.
Additionally, in that invariant mass range the contribution from QGP is expected to be dominant over hadron decays and Drell-Yang processes~\cite{Armesto_2008}.
This allows us to calculate the dilepton rate using the Born approximation~\cite{PhysRevD.35.2153,PhysRevC.89.034904}
\begin{multline} \label{eq:dilepton_rate}
    \frac{dN}{d^4 q} = \int \frac{d^3 \mathbf{k}_1}{(2\pi)^3} \frac{d^3 \mathbf{k}_2}{(2\pi)^3} f(E_1, T, \mu_{\text{B}}) f(E_2, T, \mu_{\text{B}})
\\
    \times v_{12} \sigma_{q\bar{q}}(M_{\ell \ell}^2) \delta^{(4)}(q - k_1 - k_2),
\end{multline}
where $q = (q_0,\textbf{q})$ is the four momentum, $M_{\ell \ell}$ is the dilepton invariant mass, $f(E, T, \mu_{\text{B}})$ is the Fermi-Dirac distribution for quarks and antiquarks in the QGP that includes the baryon chemical potential, $v_{12}$ is the relative velocity between a quark and an antiquark, and $\sigma_{q\bar{q}}(M_{\ell \ell}^2)$ denotes the annihilation cross section.
For a three-flavor QGP, the cross section is
\begin{equation}
\sigma_{q\bar{q}}(M_{\ell \ell}^2) = \frac{5}{9}\frac{16\pi\alpha_\text{EM}^2}{M_{\ell \ell}^2}\left( 1 + \frac{2m_\ell^2}{q^2} \right)\left( 1 - \frac{4m_\ell^2}{q^2} \right)^{1/2}.
\end{equation}
We will ignore the lepton mass $m_\ell$ and quark masses because the interested dilepton invariant mass is much larger, $M_{\ell \ell} \gg m_\mu \gg m_e \gg m_q$. 
After integrating over $\mathbf{k}_1$ and $\mathbf{k}_2$, the dilepton rate Eq.\eqref{eq:dilepton_rate} becomes,
\begin{equation}\label{eq:solved_rate}
\frac{d^4R^{\ell^+\ell^-}}{d^4q} = \frac{\alpha_{\text{EM}}^2}{6\pi^4}\frac{1}{e^{\beta q^0}-1} H(q;\beta,\mu_{\text{B}}),
\end{equation}
where
\begin{multline}
H(q;\beta,\mu_{\text{B}}) = 1 - \frac{1}{\beta |\textbf{q}|}
\ln \left[
\frac{1+e^{-\beta (q^{-}+\mu_q/2)}}
     {1+e^{-\beta (q^{+}+\mu_q/2)}}
\right]
\\
- \frac{1}{\beta |\textbf{q}|}
\ln \left[
\frac{1+e^{-\beta (q^{-}-\mu_q/2)}}
     {1+e^{-\beta (q^{+}-\mu_q/2)}}
\right],
\end{multline}
and
\begin{align}
q^\pm=\frac{q^0\pm|\textbf{q}|}{2}, && \mu_q=\frac{\mu_{\text{B}}}{3}, && \beta=\frac{1}{T}.
\end{align}

During the hydrodynamic evolution, we integrate Eq.\eqref{eq:solved_rate} over all numeric cells with an energy density larger than the freezeout energy density $e_{\text{f}} = 0.26\,\text{GeV}/\text{fm}^3$.
Because Eq.\eqref{eq:solved_rate} is written in the center of mass frame and the dilepton momentum is measured in the lab frame, we boost from the lab to the local rest frame $q^\mu = \Lambda^{\mu\nu} q'_\nu$.
This means the energy $q^0$ in Eq.\eqref{eq:solved_rate} is $q^0 = u_\mu p^\mu = \gamma \left( m_T \cosh y - v_x p_T \cos\phi - v_y p_T \sin\phi - v_z m_T \sinh y \right)$, where $u_\mu$ is the fluid four-velocity of Eqs.\eqref{eq:hydro_current}  and \eqref{eq:hydro_energymom}.

\section{RESULTS}\label{three}
We have calculated the thermal dilepton rate Eq.\eqref{eq:solved_rate} using the hydrodynamic framework of Sec.~\ref{two} at $0\%$-central Pb-Pb collisions in the LHC energy $\sqrt{s_{\text{NN}}} = 5.02$ TeV over a wide rapidity of $|y| < 7.2$.
The effect of the baryon density is evaluated by comparing the finite $n_\text{B}$ and vanishing $n_\text{B}$ results.
Lastly, the effective temperatures are extracted from the dilepton spectra, and compared against the hydrodynamic temperature to discuss their sensitivities.
\subsection{Phase diagram at distinct rapidities}
\begin{figure}[htbp]
  \centering
  \includegraphics[width=\linewidth]{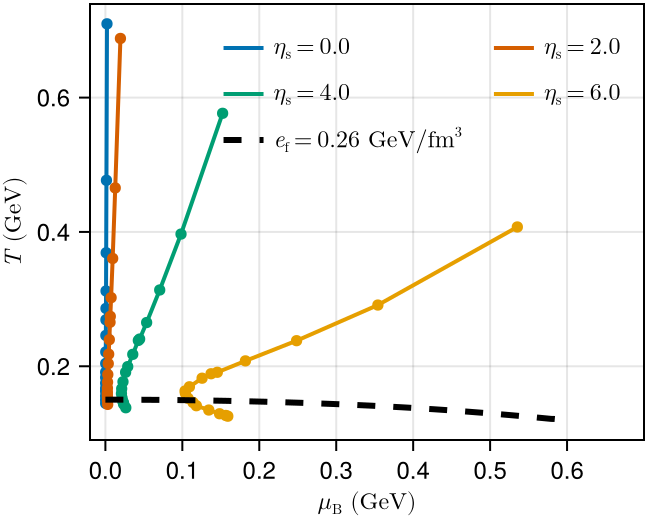}
  \caption{Temperature and baryon chemical potential of QGP in hydrodynamical calculation in QCD phase diagram. The cells at $(x,y) = (0,0)$ and $\eta_\text{s} = 0.0, 2.0, 4.0$ and $6.0$ are selected to obtain the trajectories. Dashed line represents $e = 0.26\,\text{GeV/}\text{fm}^3$ showing the chemical freezeout boundary.}
  \label{fig:trajectory}
\end{figure}

First, we present the space-time evolution of the QGP obtained from our calculations.
Figure~\ref{fig:trajectory} shows the trajectories of hydrodynamic cells on the QCD phase diagram for distinct rapidities.
The temperature $T$ and chemical potential $\mu_{\text{B}}$ were calculated at $(x,y) = (0,0)$ and different $\eta_\text{s}$.
Solid lines show the time evolution where each dot represents a step of $\Delta\tau = 1.0\text{ fm}$.
The dashed line shows the chemical freezeout boundary at $e_{\text{f}} = 0.26\,\text{GeV}/\text{fm}^3$, taken from NEOS-4D.
The cell at $\eta_\text{s} = 0$ starts at the highest temperature and lowest chemical potential because of reduced baryon stopping at high collision energies.
At forward-rapidity, the initial temperature becomes lower and the initial baryon chemical potential becomes higher.

The initial baryon chemical potential becomes a maximum at $\eta_\text{s} \sim 6$, above $\mu_{\text{B}} = 500~\text{MeV}$.
This follows from our initial conditions in Fig.~\ref{fig:baryon_distribution} and is comparable to the $\mu_{\text{B}}$ in $\sqrt{s_{\text{NN}}}=7.7~\text{GeV}$ collisions at the RHIC BES program~\cite{PhysRevC.105.034912}.
Although the temperature and baryon chemical potential decrease together through the space-time evolution of QGP, after reaching the freezeout boundary the baryon chemical potential increases.
Upon reaching the freezeout boundary, the particle composition becomes fixed.
Consequently, the temperature decreases while the baryon-to-antibaryon ratio is maintained, which causes the increase of baryon chemical potential.
This behavior arises from the transition into the hadron resonance gas model in NEOS-4D.

\subsection{Dilepton rates across space-time rapidities}
\begin{figure}[htbp]
  \centering
  \includegraphics[width=\linewidth]{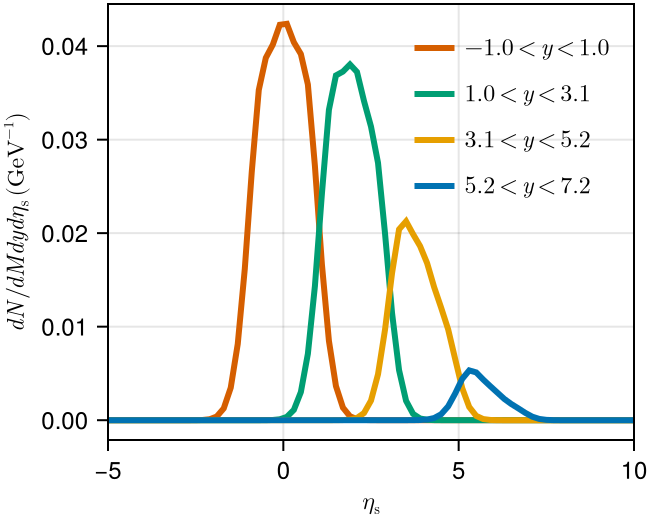}
  \caption{The space-time rapidity distributions of dilepton source for four distinct rapidity ranges.}
    \label{fig:etas_to_rapidity}
\end{figure}

Finite baryon density can suppress the thermal dilepton rate because of a decrease in available $q\bar{q}$ pairs~\cite{PhysRevLett.70.2860,PhysRevC.109.044915}.
To investigate the magnitude of that suppression, we first check a correspondence between the dilepton momentum rapidity $y$ and spatial rapidity $\eta_\text{s}$ in Fig.~\ref{fig:etas_to_rapidity}.
The two rapidities generally correspond to the same values, and the rate is inversely proportional to both the rapidities.
The center values of the four momentum rapidity regions are consistent with the spatial rapidities for the trajectories in Fig.~\ref{fig:trajectory}.
This correspondence between $y$ and $\eta_\text{s}$ is summarized in Tab.~\ref{tab:etas_rapidity}.

\begin{table}[htb]
    \centering
    \caption{Association of dilepton in specific rapidity ($y$) and space-time rapidity ($\eta_\text{s}$) of production source.}
    \label{tab:etas_rapidity}
    \begin{tabular}{|c|c|c|}
        \hline
        $\eta_\text{s}\,\text{(mid)}$ & $\eta_\text{s}\,\text{(range)}$ & $y\,\text{(range)}$ \\ \hline \hline
        0.0 & -0.2 to 0.2 & -1.0 to 1.0 \\
        \hline
        2.0 & 1.8 to 2.2 & 1.0 to 3.1 \\
        \hline
        4.0 & 3.8 to 4.2 & 3.1 to 5.2 \\
        \hline
        6.0 & 5.8 to 6.2 & 5.2 to 7.2 \\
        \hline
    \end{tabular}
\end{table}

\begin{figure}[htbp]
  \centering
  \includegraphics[width=\linewidth]{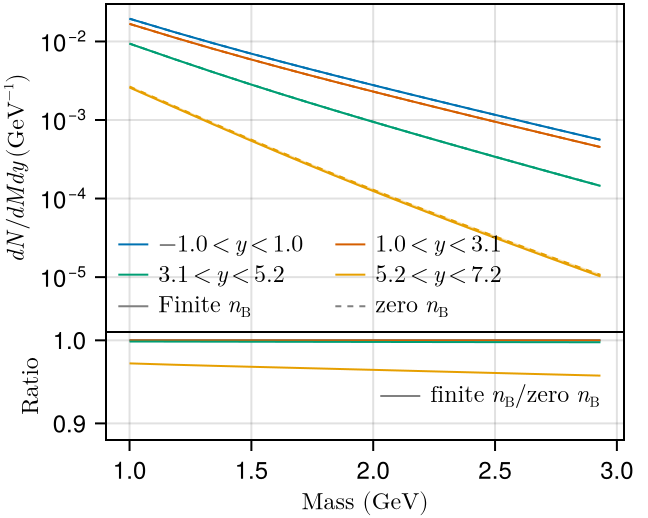}
  \caption{The rapidity dependence of thermal dilepton mass spectrum with finite and vanishing baryon density. The solid lines corresponds to the finite baryon density. The dashed line represents the zero baryon density. The bottom figure shows the ratio of the value from the finite baryon density and the zero baryon density.}
  \label{fig:integrated_thermal_dilepton_rate}
\end{figure}

Figure~\ref{fig:integrated_thermal_dilepton_rate} shows the thermal dilepton mass spectra across the rapidity regions defined in Tab.~\ref{tab:etas_rapidity}.
The results are compared with the spectra with vanishing baryon chemical potential. 
The plots are the average of four events with statistical errors indicated by shaded bands.
However, these bands are too small to be visible, as they are less than $5\%$ across all regions.
As the rapidity range increases, the rate decreases and the slope becomes steeper because of a lower source temperature.
This result is consistent with the trajectories through the phase diagram in Fig.~\ref{fig:trajectory} where the temperature decreases as the rapidity range increases.
In $y < 5.2$, the difference between the results at finite and vanishing chemical potential is below 1~\%.
By contrast, the spectra for rapidities $5.2 < y < 7.2$, where the initial $\mu_{\text{B}}$ is above $500~\text{MeV}$, show a $3-4~\%$ suppression.

We attribute the yield suppression to two reasons: a decrease in $q\bar{q}$ pairs and a slight decrease in temperature.
First, the finite $\mu_{\text{B}}$ causes the asymmetry between $q^+$ and $q^-$ that reduces the quark-antiquark annihilation processes~\cite{PhysRevLett.70.2860}.
Second the temperature decreases due to the fixed energy density with increasing $\mu_{\text{B}}$.
Both effects are significant for the finite $n_\text{B}$ result in $5.2 < y < 7.2$ leading to a sizable difference in the slope and yield.

\subsection{Effective temperature at forward rapidity}
The effective temperature of QGP can be estimated from the slope of the thermal dilepton mass spectrum~\cite{GEURTS2023104004}.
Assuming the photon production follows the Boltzmann distribution, thermal dilepton mass spectrum can be approximated as
\begin{equation}\label{eq:effective_temp}
\frac{dN}{dM_{\ell \ell}}\propto (M_{\ell \ell}T)^{3/2}\exp\left(-\frac{M_{\ell\ell}}{T}\right)
\end{equation}
where $M_{\ell\ell} \gg T$ and $M_{\ell\ell} \gg \mu_{\text{B}}$.
Equation \eqref{eq:effective_temp} relies on the assumption that the spectrum $dN/dM_{\ell\ell}$ is reference frame independent, meaning the transverse flow is small and the spectrum is boost invariant.
The correspondence of space-time rapidity and the measured rapidities for dileptons in Fig.~\ref{fig:etas_to_rapidity} and Tab.~\ref{tab:etas_rapidity} guarantee the boost invariance even in forward rapidities.
Consequently, the assumption behind Eq.~\eqref{eq:effective_temp} is satisfied, and thermal dileptons can work as a thermometer even in forward rapidities.

We extract the effective temperature $T_\text{eff}$ from the inverse slope of $\ln[M_{\ell\ell}^{3/2}(dN/dM_{\ell\ell})]$ and demonstrate the sensitivity to the true initial temperature in the forward region with finite $\mu_{\text{B}}$.
We obtained the slope from a linear least-squares method for the invariant masses $1.2 < M_{\ell\ell} < 2.6\,\text{GeV}$.
This mass range has the advantage that the radiation from QGP is dominant if open heavy flavor contributions can be removed.
The mass regions close to the $\phi~(1.02~\text{GeV})$ and $J/\psi~(3.10~\text{GeV})$ resonances are excluded from the fit mass range.

Equation~\eqref{eq:effective_temp} extracts a single effective temperature from the dilepton emission because it was integrated over spacetime.
On the other hand, in hydrodynamic calculations the temperature varies from cell to cell under the assumption of local equilibrium.
We therefore compare the effective temperature and the true temperature from our hydrodynamic calculation to test Eq.\eqref{eq:effective_temp}.
The true temperatures are taken from the cells where the energy density exceeds $e_{\text{f}}$.

\begin{figure}[htbp]
  \begin{minipage}{1.0\linewidth}
    \centering
    \includegraphics[width=\linewidth]{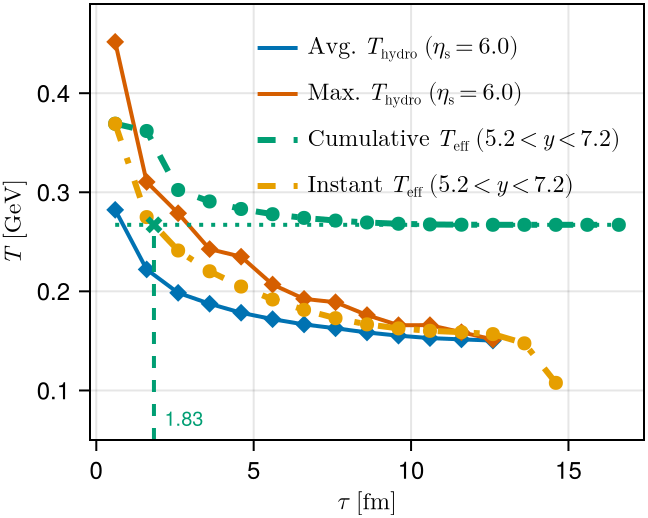}
    \subcaption{The effective temperature is obtained from dileptons in $5.2 < y < 7.2$. The maximum and average temperatures in hydrodynamical calculation is derived from the cells with $\eta_\text{s} = 6.0$.}
    \label{fig:average_temperature_instant_forward}
  \end{minipage}

  \vspace{3pt} 

  \begin{minipage}{1.0\linewidth}
    \includegraphics[width=\linewidth]{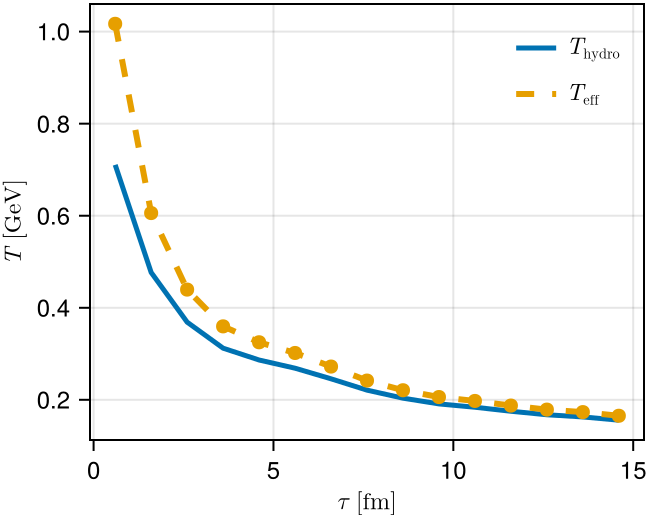}
    \subcaption{The effective temperature and the hydrodynamical temperature are obtained from single cell with $(x,y,\eta_\text{s}) = (0, 0, 0) $ for each time step.}
    \label{fig:Teff_onecell}
  \end{minipage}
  \caption{The comparison of the effective temperature and the hydrodynamical temperatures at each time steps.}
  \label{fig:average_temperatures}
\end{figure}

Figure~\ref{fig:average_temperature_instant_forward} shows the time evolution of the maximum and average hydrodynamic temperatures $T_{\text{hydro}}$ compared with the effective temperatures $T_\text{eff}$.
The cumulative effective temperatures are integrated over the spectra full spacetime, from the initial proper time $\tau_0=0.6~\text{fm}$ until each time step.
In contract, the instant effective temperatures are integrated only over space, which gives a time-dependent snapshot.

For all time slices in Fig.~\ref{fig:average_temperatures}(\subref{fig:average_temperature_instant_forward}), the instant $T_{\text{eff}}$ at a given time step is higher than the average $T_{\text{hydro}}$ and the difference is significant at earlier times due to two possible factors.
The first factor is cell-by-cell temperature fluctuation. 
The dilepton emission rate is enhanced in cells at higher temperature, due to the increased quark-antiquark density governed by the Fermi-Dirac distribution.
Because the effective temperature is integrated over multiple cells, it is strongly biased by high temperature cells, causing $T_{\text{eff}}$ to approach the maximum $T_{\text{hydro}}$. 
In later times, as the difference of the average $T_{\text{hydro}}$ and the maximum $T_{\text{hydro}}$ narrows, the instant $T_{\text{eff}}$ approaches these temperatures.
The second factor is the approximation of $T \ll M_{\ell\ell}$ that tends to deviate at high temperature.
Figure \ref{fig:average_temperatures}(\subref{fig:Teff_onecell}) demonstrates the deviation of the effective temperature from the hydrodynamic temperature in a cell at $(x,y,\eta_\text{s})=(0,0,0)$.
The deviation is more significant with high temperatures in an early time, while they agree reasonably well in the low temperature region.
These two factors cause the discrepancy toward higher temperature. 

The dilepton spectra available in experiments represent the cumulative effective temperature over the full time evolution.
The cumulative effective temperature in Fig. \ref{fig:average_temperatures}(\subref{fig:average_temperature_instant_forward}) decreases significantly in the beginning, but becomes almost constant after $\tau= 10.0\,\text{fm}$.
This behavior originates from predominant contributions in the early stage with high temperature.
The cumulative effective temperature at last is $0.27\,\text{GeV}$.
It corresponds to the instant effective temperature at $\tau = 1.83\,\text{fm}$, which can be regarded as the central time probed by the measurable effective temperature.



\begin{figure}[htbp]
  \centering
  \includegraphics[width=\linewidth]{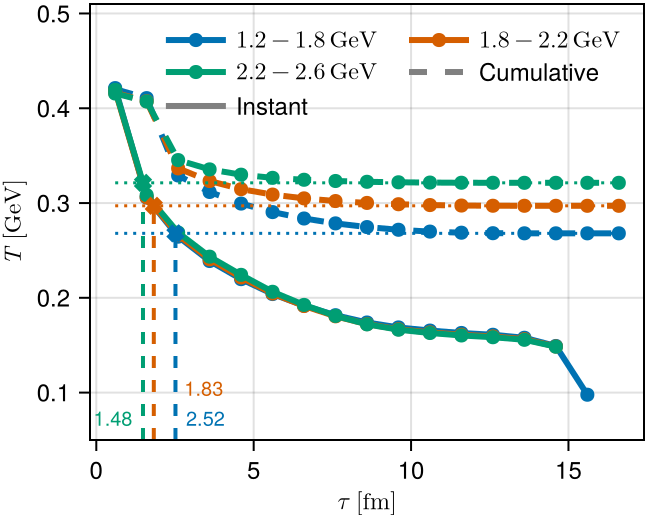}
  \caption{Fit mass range dependence of the effective temperatures in $-5.2 < y < 7.2$. The mass range is divided into three regions in $1.2 - 2.6~\text{GeV}$. The solid line shows the results from cumulative dileptons from initial state, and the dashed line shows the results from each time steps.}
  \label{fig:average_temperature_cumulative_fwd_four}
\end{figure}

Since a higher temperature causes a shallower slope of the mass spectrum, it is expected that the contribution from higher temperature region increases in the higher mass region.
Therefore, the time scale accessible by the effective temperature depends on the mass region used for the fitting, as it is confirmed in Ref.~\cite{physjc85}.
Here, the fitting mass range dependence of the effective temperature is investigated in forward rapidity regions as shown in Fig. \ref{fig:average_temperature_cumulative_fwd_four}.
The mass range is divided into three regions in $1.2 < M_{\ell\ell} < 2.6\,\text{GeV}$.
The proper time of instant effective temperature corresponding to the cumulative effective temperature for each fitting region is summarized in Tab. \ref{tab:mass_range_fwd}.
The correspondence between the probing proper time and the effective temperatures shows a clear dependence on fitting mass range; a temperature in an earlier proper time is probed in a higher mass region.

\begin{table}[htb]
    \centering
    \caption{The proper time of instant effective temperature corresponding to the cumulative effective temperature for each fitting region in $5.2 < y < 7.2$.}
    \label{tab:mass_range_fwd}
  \begin{tabular}{|c||c|c|c|}
        \hline
        \text{Fitting mass region (GeV)} & 1.2 - 1.8 & 1.8 - 2.2 & 2.2 - 2.6  \\
        \hline
        \text{Proper time} & $2.52\,\text{fm}$ & $1.83\,\text{fm}$ &  $1.48\,\text{fm}$ \\
        \hline
  \end{tabular}
\end{table}

Finally, we discuss the possibility of thermal dileptons as a baryometer.
Probing $\mu_{\text{B}}$ requires the detection of the thermal dilepton suppression as shown in Fig.~\ref{fig:integrated_thermal_dilepton_rate}. 
To achieve this, the dilepton yield with vanishing $\mu_{\text{B}}$ needs to serve as a baseline for the comparison, which can be accomplished by the parametrization of dilepton yield in vanishing $\mu_{\text{B}}$ in terms of $T_{\text{eff}}$ and $y$.
$T_{\text{eff}}$ has no dependence of $\mu_{\text{B}}$ as Eq.~\eqref{eq:effective_temp} and holds the validity as a thermometer under the assumption of the boost invariance of the expansion in $\eta_\text{s}$ direction.
The yield with vanishing $\mu_{\text{B}}$ can be normalized by the spectra in the mid rapidity region where $\mu_{\text{B}} \sim 0$.
The challenge lies in parametrization of the $y$ dependence of the yield arising from factors other than $T$ and $\mu_{\text{B}}$. 
Once this parametrization is established from theoretical model or observables, it will enables to probe $\mu_{\text{B}}$ by comparing the predicted value with measured yield.

\section{CONCLUSION}\label{four}
We have demonstrated the impact of the finite $\mu_{\text{B}}$ on the thermal dilepton mass spectrum in heavy-ion collisions at the LHC energy.
The space-time evolution of QGP was simulated by the $(3+1)$D relativistic ideal hydrodynamics, modified to incorporate the finite baryon density.
The initial energy density was tuned to reproduced the charged particle multiplicity measured by the ALICE experiment, and the initial baryon density was determined to get as close as possible to the McDIPPER results.
Our results indicated the large $\mu_{\text{B}}$, which is comparable to that achieved in $\sqrt{s_{\text{NN}}} = 7.7~\text{GeV}$ in the BES program calculated by AMTP model~\cite{PhysRevC.105.034912}, can be reached in the forward region that support the findings in Ref.~\cite{PhysRevC.99.014906}.
The dilepton mass spectrum in $1 < M_{\ell\ell} < 3\text{ GeV}$ was calculated across a wide rapidity range.
We found that the finite $\mu_{\text{B}}$ suppresses the dilepton rate for 3--4$~\%$ in $5.2 < y < 7.2$, while the suppression is below $1~\%$ in $y < 5.2$.

In addition, the effective temperature extracted from the spectra were compared with the true temperature in the hydrodynamic calculation.
The effective temperature was found to be consistently larger than the true temperature but still correlates strongly with the initial QGP temperature in the forward rapidity region with a large $\mu_{\text{B}}$.
The proper time probed by this correlation depends on the invariant mass range choice for the fit.

In the present study, we employed ideal hydrodynamics to isolate the impact of finite baryon density on thermal dilepton 
production in the forward rapidity region. While viscous effects are known to influence the space-time evolution of 
the medium and dilepton emission rates, the primary goal of this work is to establish the magnitude of the finite-$\mu_B$ effect 
in a previously unexplored kinematic regime. A quantitative assessment including shear and baryon diffusion effects will be important for future studies.

Our study suggests that the forward-rapidity region of Pb--Pb collisions at the LHC provides 
access to a thermodynamic regime that is distinct from the nearly baryon-free matter commonly studied at midrapidity. 
In this regime, temperatures characteristic of the LHC coexist with baryon chemical potentials comparable to those 
achieved in the RHIC BES program. 
Thermal dileptons remain sensitive to this medium while preserving their role as effective thermometers. 
These features make forward-rapidity dileptons a promising tool for exploring the properties of high-temperature finite-density QCD matter.

Although the predicted suppression of the dilepton yield remains modest, our results indicate 
potentially observable with future forward dilepton measurements. 
Together with future improvements in detector coverage and statistical precision, thermal dileptons 
may provide not only a thermometer of the QGP but also a complementary probe of baryon transport dynamics at the LHC.
These findings motivate future thermal dilepton measurements over an extended rapidity range at the LHC. 
Such measurements would provide a unique opportunity to investigate a finite-density QGP, 
test predictions of longitudinal baryon transport, and further establish thermal dileptons as probes of both temperature and baryon density 
in high-energy heavy-ion collisions.


\begin{acknowledgments}
We thank A. Monnai for discussion about the implementation of equation-of-state. The numerical calculation in this study was carried out at the Yukawa Institute Computer Facility.
This work was also supported by JSPS KAKENHI Grant Numbers, JP20H00156, JP20H11581, JP2500449 (C.N.) and by the World Premier International Research Center Initiative (WPI) under MEXT, Japan (C.N.).
\end{acknowledgments}

\appendix
\section{Effective temperature at mid rapidity}
\begin{figure}[htbp]
  \centering
  \includegraphics[width=\linewidth]{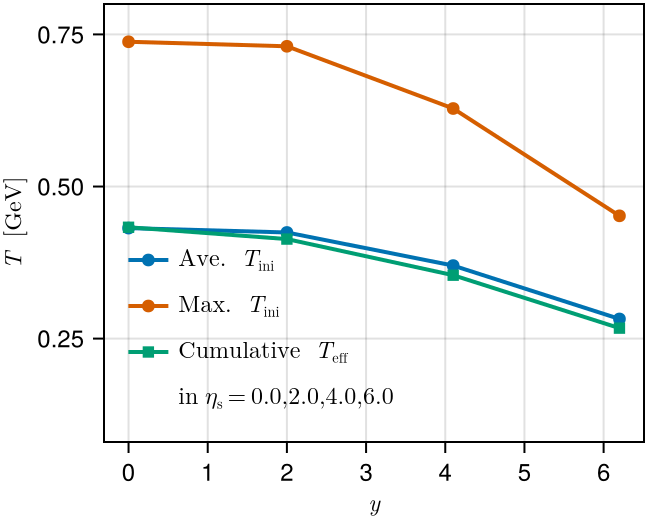}
  \caption{Rapidity distribution of the effective temperatures and initial/maximum temperatures from hydrodynamical calculation. The correspondence of $y$ and $\eta_\text{s} $ follows Tab.~\ref{tab:etas_rapidity}.}
  \label{fig:y_t}
\end{figure}

This appendix provides a comparison between the effective and true temperatures in the mid rapidity region.
While the sensitivity at mid rapidity has been well studied, this comparison serves as a benchmark for our main analysis, which focuses on the forward rapidity region.
Figure~\ref{fig:y_t} shows the rapidity dependence of the effective temperature, compared with the initial temperature in hydrodynamical calculations.
It is based on the mapping between the dilepton $y$ and $\eta_\text{s}$ summarized in Tab.~\ref{tab:etas_rapidity}.  
The effective temperatures closely reflect the initial temperatures, showing a decrease with increasing rapidity.
While the effective temperatures are naively expected to be lower than the initial temperature due to the accumulation of radiation from later stages with lower temperature, this effect is offset by the upward deviation in the effective temperature estimation as discussed in Fig.~\ref{fig:average_temperature_instant_forward}. 
Consequently, the cumulative effective temperatures are close to the initial average temperature.

\begin{figure}[htbp]
  \centering
  \includegraphics[width=\linewidth]{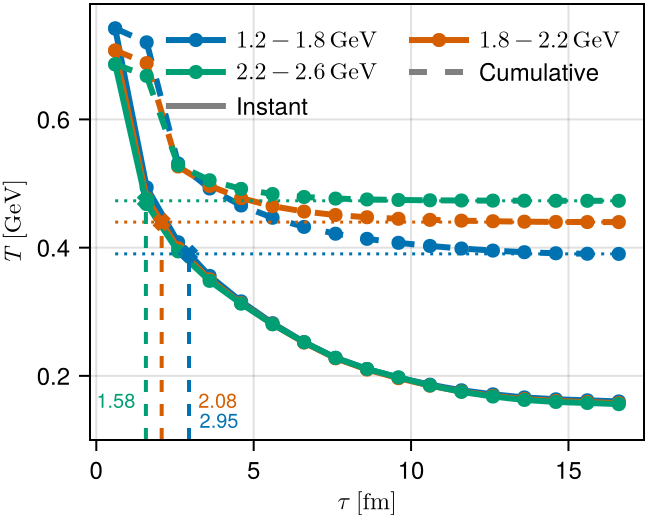}
  \caption{Fit mass range dependence of the effective temperatures in $-1.0 < y < 1.0$. The mass range is divided into three regions in $1.2 - 2.6~\text{GeV}$. The solid line shows the results from cumulative dileptons from initial state, and the dashed line shows the results from each time steps.}
  \label{fig:average_temperature_cumulative_mid_four}
\end{figure}

Figure~\ref{fig:average_temperature_cumulative_mid_four} shows the dependence of the effective temperature on the fit mass range within the mid rapidity region ($-1.0 < y < 1.0$).
The results are similar to those obtained for the forward rapidity in Fig.~\ref{fig:average_temperature_cumulative_fwd_four}, indicating that the sensitive time scale exhibits a clear mass dependence.
The proper time for each fitting region is summarized in Tab. \ref{tab:mass_range_mid}.

\begin{table}[htb]
    \centering
    \caption{The proper time of instant effective temperature corresponding to the cumulative effective temperature for each fitting region in $-1.0 < y < 1.0$.}
    \label{tab:mass_range_mid}
  \begin{tabular}{|c||c|c|c|}
        \hline
        \text{Fitting mass region (GeV)} & 1.2 - 1.8 & 1.8 - 2.2 & 2.2 - 2.6  \\
        \hline
        \text{Proper time} & $2.95\,\text{fm}$ & $2.08\,\text{fm}$ &  $1.58\,\text{fm}$ \\
        \hline
  \end{tabular}
\end{table}
\bibliography{apssamp}

\end{document}